\documentclass[fleqn,12pt,twoside]{article}
\usepackage[headings]{espcrc1}

\readRCS
$Id: espcrc1.tex,v 1.2 2004/02/24 11:22:11 spepping Exp $
\usepackage{graphicx}
\usepackage[figuresright]{rotating}

\newcommand{\AmS}{{\protect\the\textfont2
  A\kern-.1667em\lower.5ex\hbox{M}\kern-.125emS}}

\title{Jet and Leading Hadron Production in High-energy Heavy-ion Collisions}

\author{Xin-Nian Wang\address{Nuclear Science Division, MS R0319 \\
    Lawrence Berkeley National Laboratory, Berkeley, CA 94720}}
       
\runtitle{Jet and leading hadron production}
\runauthor{X.-N. Wang}

\begin{document}

%LBNL-59008
% typeset front matter
\maketitle

\begin{abstract}
Jet tomography has become a powerful tool for the study of properties of 
dense matter in high-energy heavy-ion collisions. I will discuss recent
progresses in the phenomenological study of jet quenching, including
momentum, colliding energy and nuclear size dependence of single hadron
suppression, modification of dihadron correlations and the soft hadron
distribution associated with a quenched jet.
\end{abstract}

\section{Introduction}

The most important consequence of the discovery of the asymptotic freedom
of QCD \cite{asymp} is the small value of strong coupling constant $\alpha_{\rm s}$ 
at short distances or in hard processes involving large energy and momentum
transfer. It makes the perturbative expansion in $\alpha_{\rm s}$
a reliable technique for calculations of many physical observables
in these hard processes. Hadronic interaction and production often 
involve strong interaction at long distance, which is not calculable 
within the framework of the perturbative expansion. However, it has been 
proven to the leading power correction ($1/Q^2$) that the cross 
section can be factorized into short-distance parts calculable in 
perturbative QCD (pQCD) and non-perturbative long distance parts \cite{collin}. 
These long distance parts can be expressed as matrix elements that are 
universal. Therefore, they can be measured in one process and used in 
another process; therein lies the predictive power of pQCD. It
has been extremely successful in the study of hard processes, from
electron-positron annihilation, to deeply inelastic scattering (DIS)
of leptons and nucleons, Drell-Yan dilepton production and large 
transverse momentum jet production in hadronic collisions.

Hard processes can also happen in high-energy nucleus-nucleus collisions.
They include production of large transverse momentum jets, direct photons 
and dileptons with large invariant-mass and heavy quarks. Since hard 
processes happen on a short time scale in the earliest stage of 
high-energy heavy-ion collisions, they can probe the bulk matter that 
is formed shortly after the collision. The pQCD parton model serves as a 
reliable and tested framework for the study of these hard probes. 
In this talk, I will focus on the physics of jet propagation in the 
dense medium and recent progresses in the phenomenological study of
jet and high $p_T$ hadron production in heavy-ion collisions.

The study of jet production in heavy-ion collisions exploits the attenuation 
of parton jets or jet quenching during their propagation in dense 
medium. Such an idea was first proposed by Bjorken \cite{bj}
to study the space-time structure of high-energy
hadron-hadron collisions via elastic  energy loss of partons
in medium \cite{bj}. But it was soon realized that elastic energy
loss may be very small relative to the radiative energy loss induced by
multiple scatterings \cite{gp90}. The effect of jet attenuation
in medium and its utilization as a probe was taken more seriously
only after a Monte Carlo study within the HIJING model \cite{gw92} that 
demonstrated
significant suppression of high $p_T$ single inclusive hadron spectra
due to jet quenching. Such large suppression can be easily measured in
experiments without jet reconstruction and detailed study of medium
modification of jet structure. It is quite amazing how well the
calculated high $p_T$ hadron suppression with a very crude
estimate (or rather a guess) and simulation of average parton energy loss,
$dE/dx=1$ GeV/fm, agrees qualitatively with the first RHIC data
in central $Au+Au$ collisions \cite{Adcox:2001jp}. However, later
theoretical studies show that radiative parton energy loss involves
quantum field treatment of induced radiation and the non-Abelian
Landau-Pomeranchuk-Migdal interference effects in QCD \cite{eloss}.
The non-Abelian energy loss depends on both the local parton density
and the total distance of the parton propagation.

It was realized from the beginning that one has to understand the
cold-nuclear effects on high $p_T$ hadron spectra in $p+A$ collisions in
order to extract the genuine suppression caused by the hot medium in heavy-ion
collisions \cite{gw92}. These cold-nuclear effects include nuclear modification
of the parton distributions in nuclei and change of high $p_T$ hadron
spectra caused by initial and final state multiple scattering in cold nuclei.
Nuclear shadowing or depletion of effective parton distributions in nuclei
is limited to small fractional momentum $x<0.1$ and thus to low $p_T$
hadron spectra $p_T<2\sim 4$ GeV/$c$ at $\sqrt{s}=200$ GeV \cite{gw92}. 
Multiple initial parton scattering and the consequent intrinsic 
transverse momentum broadening
will partially compensate the effect of nuclear shadowing. The final
hadron spectra in $p+A$ collisions at mid-rapidity were predicted to 
be enhanced slightly, known as the Cronin effect, at intermediate $p_T$ 
and then become nearly identical to that in $p+p$ collisions at 
large $p_T$ \cite{Wang:1998ww}. These predictions were indeed
verified by experimental data in $d+Au$ collisions at RHIC \cite{dau}. 
Therefore, the final state interaction and jet quenching have been 
established as the true cause of the 
observed suppression of high $p_T$ hadron spectra in $Au+Au$ 
collisions. Since parton energy loss depends on both the local parton
density and total propagation length, it will depend on the azimuthal 
angle of the jet propagation relative to the reaction plane in non-central
heavy-ion collisions. Such azimuthal angle dependence was predicted \cite{wangv2} 
to give rise to an azimuthal anisotropy of the suppressed large $p_T$
hadron spectra which was also observed at RHIC \cite{starv2}. The most
striking consequence of jet quenching observed so far is the suppression
of the away-side jets in two-hadron correlation measurements \cite{star-jet},
providing a clear illustration of the jet quenching picture in heavy-ion
collisions.

\section{Jet quenching and modified jet fragmentation}

Though jet quenching can be intuitively related to parton energy loss,
the experimentally measurable consequences can only be found in the
modification of the final hadron spectra from jet fragmentation in
medium relative to that in vacuum. In addition, there are many partonic
processes, such as quark-anti-quark annihilation, in which the 
leading parton can lose their identities (flavor or gluon versus quark).
In these processes, the concept of parton energy loss become very
ambiguous.  Furthermore, hard processes are normally followed by final state 
radiations with a short time even in the vacuum. The DGLAP evolution 
and induced radiation due to final state multiple scattering should 
be considered together in the same framework of jet fragmentation.

Within a framework of twist-expansion in the collinear factorized
parton model, one can study such medium modification of the jet
fragmentation function via modified DGLAP evolution equations \cite{wgdis}.
For quark propagation in a cold nuclear medium in deeply inelastic
lepton scattering off a nuclear target, the modified fragmentation
function,
\begin{eqnarray}
\widetilde{D}_{q\rightarrow h}(z_h,Q^2)&\equiv&
D_{q\rightarrow h}(z_h,Q^2)
+\int_0^{Q^2} \frac{d\ell_T^2}{\ell_T^2}
\frac{\alpha_s}{2\pi} \int_{z_h}^1 \frac{dz}{z}
\left[ \Delta\gamma_{q\rightarrow qg}(z,x,x_L,\ell_T^2)\right.
 D_{q\rightarrow h}(z_h/z)  \nonumber \\
&+& \left. \Delta\gamma_{q\rightarrow qg}(1-z,x,x_L,\ell_T^2)
D_{g\rightarrow h}(z_h/z)\right] \, , \label{eq:MDq}
\end{eqnarray}
has the same form as the DGLAP correction in vacuum, which
determines the evolution of $D_{q\rightarrow h}(z_h,Q^2)$ and
$D_{g\rightarrow h}(z_h,Q^2)$ as the leading-twist
quark and gluon fragmentation functions. The difference from
the vacuum DGLAP evolution lies in the modified splitting functions
\begin{eqnarray}
\Delta\gamma_{q\rightarrow qg}(z,x,x_L,\ell_T^2)&=&
\frac{1+z^2}{(1-z)_+}T^{A}_{qg}(x,x_L)
\frac{2\pi\alpha_s C_A}
{(\ell_T^2+\langle k_T^2\rangle) N_c f_q^A(x,\mu_I^2)}
+({\rm virtual\,\, corr.}) \, ,
\label{eq:r1}
\end{eqnarray}
which depends on the properties of the medium through the twist-four 
parton matrix element of the nucleus, 
\begin{eqnarray}
T^{A}_{qg}(x,x_L)&=& \int \frac{dy^{-}}{2\pi}\, dy_1^-dy_2^-
e^{i(x+x_L)p^+y^-}(1-e^{-ix_Lp^+y_2^-})(1-e^{-ix_Lp^+(y^--y_1^-)})
\nonumber  \\
&&\times\frac{1}{2}\langle A | \bar{\psi}_q(0)\,
\gamma^+\, F_{\sigma}^{\ +}(y_{2}^{-})\, F^{+\sigma}(y_1^{-})\,\psi_q(y^{-})
| A\rangle  \theta(-y_2^-)\theta(y^- -y_1^-)
\;\; . \label{Tqg}
\end{eqnarray}
Here, the fractional momentum $x=x_B$ is carried by the initial 
quark and $x_L =\ell_T^2/2p^+q^-z(1-z)$ is the additional momentum
fraction required for induced gluon radiation.
The dipole-like structure in the above twist-four 
parton matrix element is a result of the LPM interference 
in gluon bremsstrahlung. Assuming a factorized form of the
twist-four matrix element, the above modified fragmentation function
can describe the suppression of leading hadron spectra in
HERMES data of DIS \cite{ww02} very well, including the quadratic
nuclear size dependence.

One can quantify the modification of the fragmentation
by the quark energy loss which can be defined as the momentum fraction
carried by the radiated gluon,
\begin{eqnarray}
\langle\Delta z_g\rangle
&=& \int_0^{Q^2}\frac{d\ell_T^2}{\ell_T^2}
\int_0^1 dz \frac{\alpha_s}{2\pi}
 z\,\Delta\gamma_{q\rightarrow gq}(z,x_B,x_L,\ell_T^2).
\label{eq:heli-loss}
\end{eqnarray}

To extend the study of modified fragmentation functions to 
jets in heavy-ion collisions, one can
assume $\langle k_T^2\rangle\approx \mu^2$ (the Debye screening mass)
and a gluon density profile
$\rho(y)=(\tau_0/\tau)\theta(R_A-y)\rho_0$ for a 1-dimensional 
expanding system. Since the initial jet production 
rate is independent of the final gluon density, which can be 
related to the parton-gluon scattering cross section 
[$\alpha_s x_TG(x_T)\simeq (N_c/2\pi^2) \mu^2\sigma_g$], one has then
\begin{equation}
\frac{\alpha_s T_{qg}^A(x_B,x_L)}{f_q^A(x_B)} \approx \frac{N_c}{\pi}
\mu^2\int dy \sigma _g \rho(y)
[1-\cos(y/\tau_f)],
\end{equation}
where $\tau_f=2Ez(1-z)/\ell_T^2$ is the gluon formation time. In the
limit of high initial jet energy $E$, the total energy loss becomes \cite{ww02}
\begin{equation}
\langle \Delta E \rangle \approx \pi C_aC_A\alpha_s^3
\int_{\tau_0}^{R_A} d\tau (\tau-\tau_0) 
\rho(\tau,\vec{r}_0+\vec{n}(\tau-\tau_0)) \ln\frac{2E}{\tau\mu^2},
\label{effloss}
\end{equation}
where $\sigma_g\approx C_a 2\pi\alpha_s^2/\mu^2$ ($C_a$=1 for $qg$ and 9/4 for
$gg$ scattering) is assumed. Taking into account the kinematic limits
in the integration, the total energy loss has a strong energy dependence
for finite values of $E$ \cite{glv}.

\section{Jet quenching in heavy-ion collisions}

Working in the same framework of twist expansion in the collinear
factorized parton model, one can similarly obtain the single inclusive
hadron spectra at high $p_T$ \cite{Wang:1998ww},
\begin{eqnarray}
  \frac{d\sigma^h_{AA}}{dyd^2p_T}&=&K\sum_{abcd} 
  \int d^2b d^2r dx_a dx_b d^2k_{aT} d^2k_{bT}
  t_A(r)t_A(|{\bf b}-{\bf r}|) 
  g_A(k_{aT},r)  g_A(k_{bT},|{\bf b}-{\bf r}|) \nonumber \\
  &\times& f_{a/A}(x_a,Q^2,r)f_{b/A}(x_b,Q^2,|{\bf b}-{\bf r}|)
  \frac{d\sigma}{d\hat{t}}(ab\rightarrow cd)
 \frac{D_{h/c}^\prime (z_c,Q^2,\Delta E_c)}{\pi z_c}, \label{eq:nch_AA}
\end{eqnarray}
with medium modified fragmentation functions $\widetilde D_{h/c}$.
Here, $z_c=p_T/p_{Tc}$, $y=y_c$, $\sigma(ab\rightarrow cd)$ are 
elementary parton scattering cross sections and $t_A(b)$ is the 
nuclear thickness function normalized to $\int d^2b t_A(b)=A$.
The $K\approx 1.5$--2 factor is used to account for higher order pQCD 
corrections. For simplification, many studies assume the modified 
fragmentation functions as given by the vacuum ones with the
fractional momentum rescaled by $1/(1-\Delta z)$. Such effective
description is found to reproduce 
the full calculation very well, but only when $\Delta z=\Delta E_c/E$ 
is set to be $\Delta z\approx 0.6 \langle z_g\rangle$.
One can similarly calculate back-to-back dihadron spectra in the
same parton model, assuming two jets fragment independently as described
by medium modified fragmentation functions \cite{Wang:2003mm}.
Such a parton model with modified fragmentation functions
is the framework for many phenomenological studies of jet quenching
in heavy-ion collisions \cite{Wang:1998ww,models}. It can describe well 
the suppression of the single hadron spectra, back-to-back hadron 
correlation and azimuthal angle anisotropy \cite{Wang:2003mm}, which are 
three different consequences of jet quenching. Since jet quenching
depends on the initial gluon density of the hot medium which is not
known within the parton model, one has to fit the data on the
suppression of single hadron spectra in the most central $Au+Au$ collisions
at a given colliding energy, {\it e.g.}, $\sqrt{s}=200$ GeV. Then the
centrality dependence, the energy dependence, the back-side suppression
and the azimuthal anisotropy are all predictions within the parton
model.  Combining measurements of the above three different effects
of jet quenching and compare with the results from jet quenching in
deeply inelastic $e+A$ collisions, one can conclude that the
initial gluon density (at initial time $\tau_0=0.2$ fm/$c$) reached in 
central $Au+Au$ collisions at $\sqrt{s}=200$ GeV is about 30 times 
higher than in a cold $Au$ nuclei \cite{Wang:2003mm,ww02}.

\begin{figure}[htb]
\begin{minipage}[t]{75mm}
\includegraphics[scale=0.44]{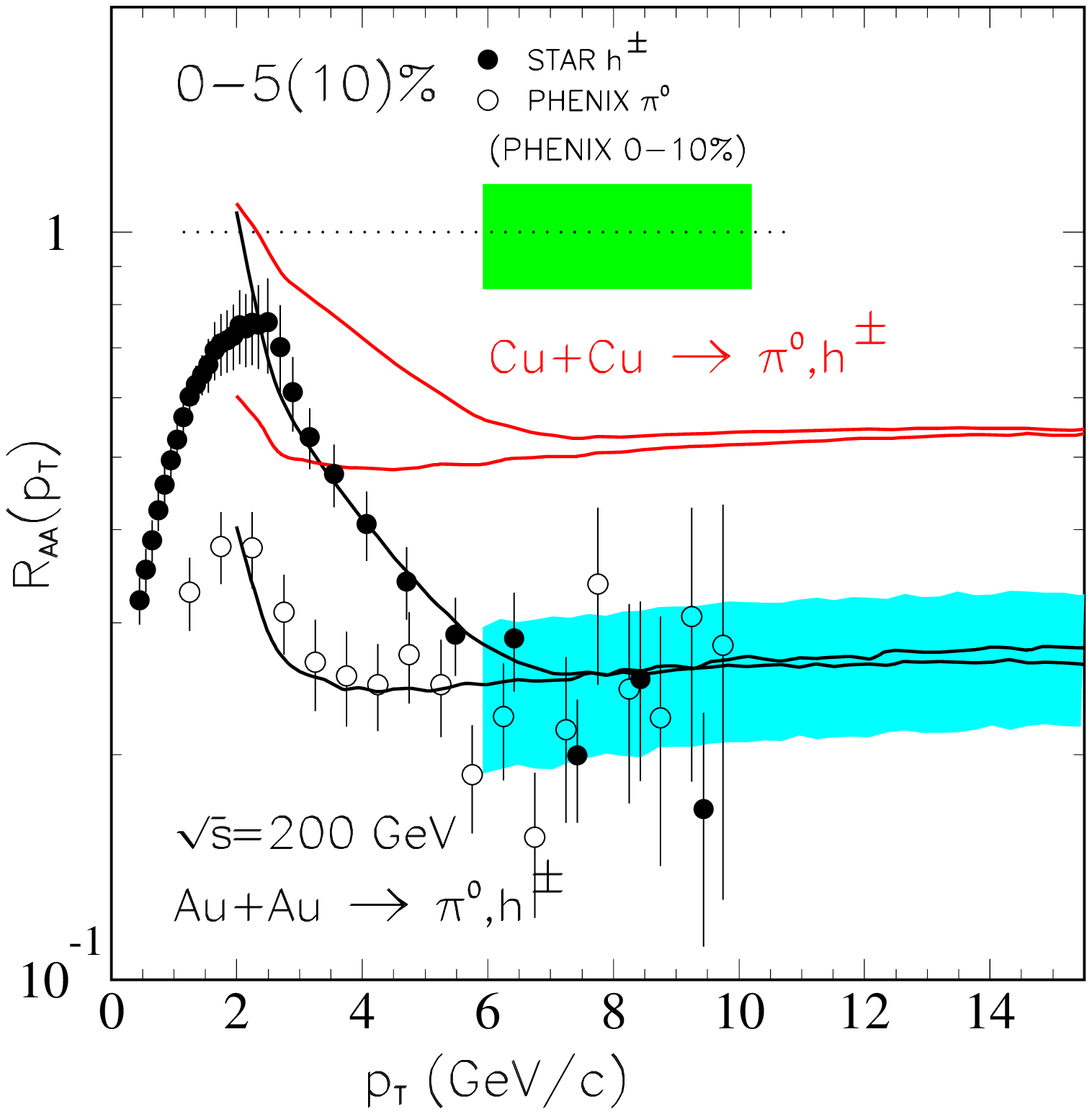}
\caption{Hadron suppression factor $R_{AA}(p_T)$ for the most central
(0-10\%) $Au+Au$ and $Cu+Cu$ collisions at $\sqrt{s}=200$ GeV.}
\label{fig-raa}
\end{minipage}
\hspace{\fill}
\begin{minipage}[t]{80mm}
\includegraphics[scale=0.5]{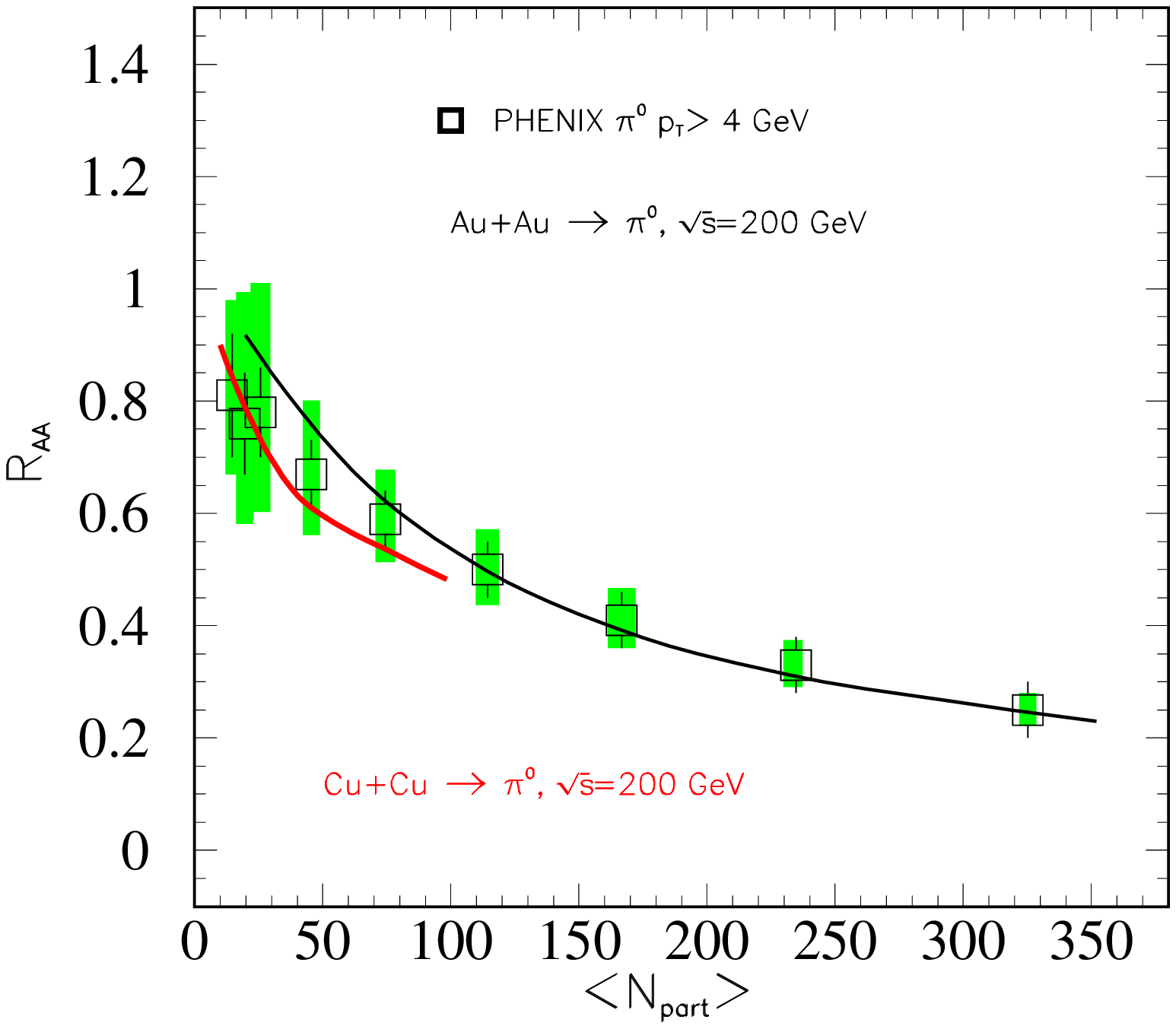}
\caption{The hadron suppression factor $R_{AA}(p_T)$ at fixed $p_T$ 
and $\sqrt{s}=200$ GeV as a function of $N_{\rm part}$ in $Au+Au$
and $Cu+Cu$ collisions.}
\label{fig-bdep}
\end{minipage}
\end{figure}

Since parton degradation in medium depends both on the local gluon
density and the propagation length (which is non-linear due to non-Abelian
LPM interference effect), the centrality dependence of the single
inclusive hadron suppression reflects a combination of these two dependencies.
One can extend the study of the density and length dependence by
varying the nuclear size at a fixed energy. Shown in Fig.~\ref{fig-raa}
are the parton model calculations of the suppression factor $R_{AA}(p_T)$
for the most 0-10\% central $Cu+Cu$ collisions at $\sqrt{s}=200$ GeV
together with the calculation and experimental data of central $Au+Au$
collisions. As expected, the suppression is very similar to
semi-peripheral (30-40\%) $Au+Au$ collisions with the same $p_T$ 
dependence. This is exactly what is observed by the experiments on
$Cu+Cu$ collisions \cite{starqm05,phenixqm05}. In principle, the suppression
factor should only be a function of the total energy loss. At fixed
energy and $p_T$, it is proportional to the path integral in 
Eq.~(\ref{effloss}). As shown in Fig.~\ref{fig-bdep} the suppression
factor at fixed $p_T\approx 4$ GeV and energy $\sqrt{s}=200$ GeV
is only approximately a function of $N_{\rm part}$. The new data
on $Cu+Cu$ collisions indeed indicate small but finite deviation 
from $N_{\rm part}$ scaling \cite{starqm05,phenixqm05}. However, the
statistical and most importantly the systematic errors are still
too big to quantify the small deviation.

\begin{figure}[htb]
\begin{minipage}[b]{72mm}
\includegraphics[scale=0.40]{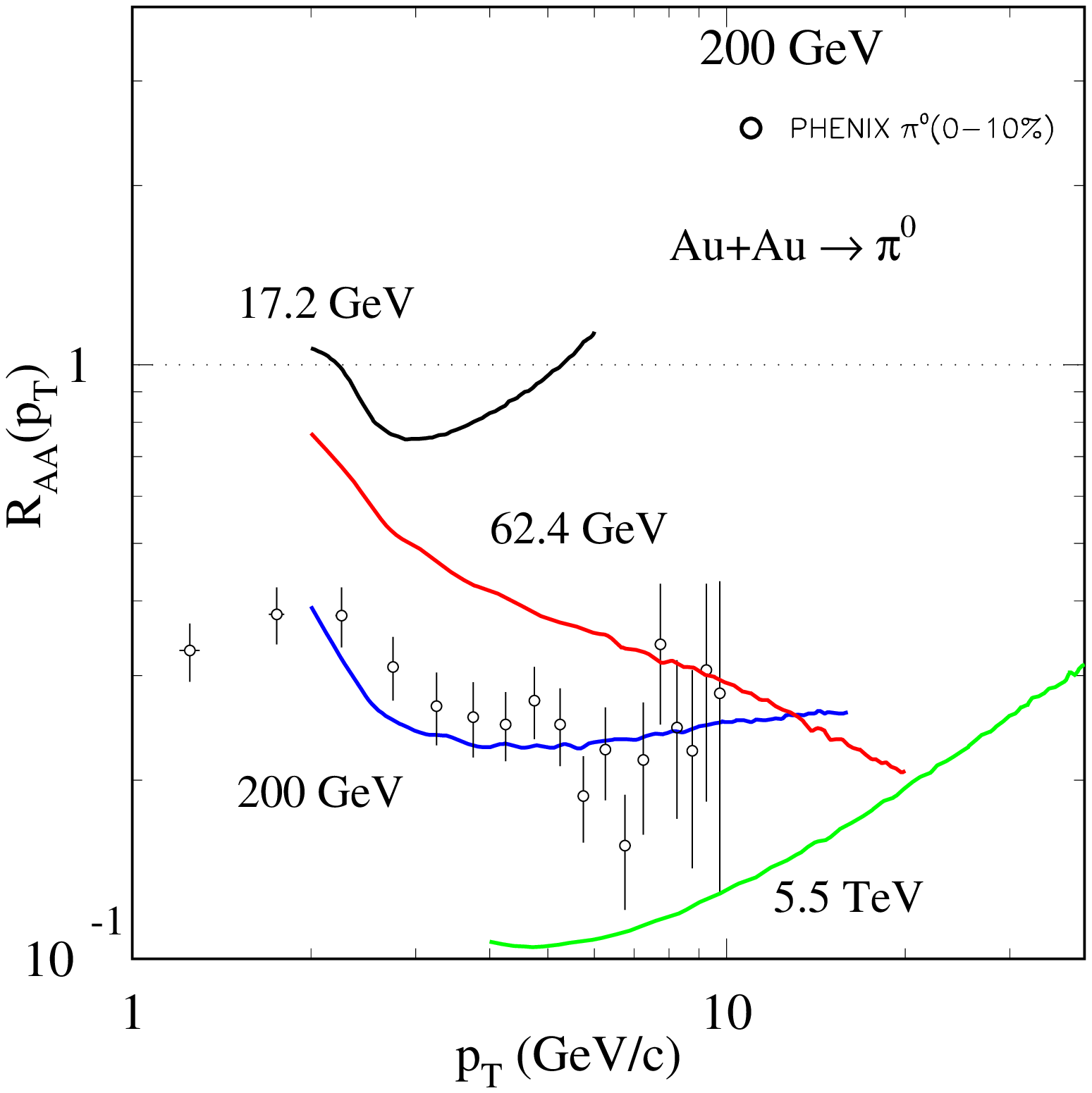}
\caption{Hadron suppression factor $R_{AA}(p_T)$ for the most central
(0-10\%) $Au+Au$ collisions at different colliding energies.}
\label{fig-raa2}
\end{minipage}
\hspace{\fill}
\begin{minipage}[b]{83mm}
\includegraphics[scale=1.07]{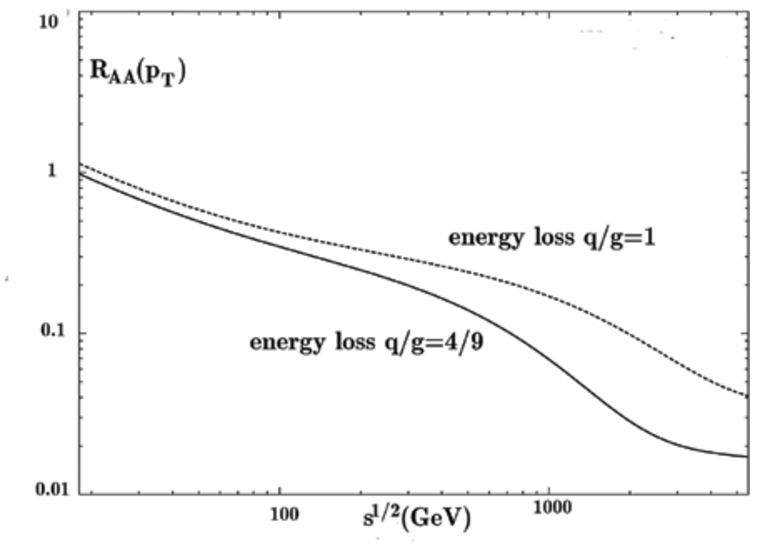}
\caption{Nuclear modification factor $R_{AuAu}$ for neutral pions
as function of collision energy at fixed $p_T=6$ GeV in central
collisions (0-10\%) with both the QCD and a non-QCD energy loss.}
\label{fig-edep}
\end{minipage}
\end{figure}

The $p_T$ dependence of the hadron suppression factor is sensitive
to many aspects of the jet production and parton energy loss.
Because of trigger bias, the initial jet energy for fixed hadron
$p_T$ varies with colliding energy $\sqrt{s}$ because the shape of
the jet spectra change dramatically from the SPS to RHIC and LHC
energies. At the same time, the parton energy loss has also a strong
energy dependence on the initial jet energy in the kinematic
regime of current experimental data. For given amount of parton energy
loss, the hadron suppression factor is also sensitive to the slope of 
the initial jet spectra which changes with $x_T=2p_T/\sqrt{s}$. In
the kinematic regime $p_T\sim 10$ GeV, jet spectra at low energies are 
very steep at the edge of kinematic limit. Any given parton energy loss 
will lead to the increase of suppression at larger $p_T$. In this 
case $R_{AA}(p_T)$, shown in Fig.~\ref{fig-raa2}, decreases with $p_T$ 
as observed in experiments
at $\sqrt{s}=63$ GeV \cite{starqm05,phenixqm05}. At $\sqrt{s}=200$ GeV, 
such effect due to the shape of jet spectra is compensated by the energy
dependence of the parton energy loss. The resulting $R_{AA}(p_T)$ is
almost independent of $p_T$ in the range of $p_T=5-20$ GeV. At
LHC energy, $p_T \sim 50$ GeV is far away from the kinematic limit.
Jet spectra have a nice power-law behavior. In this case, the 
suppression factor increases slowly with $p_T$. At the SPS energy,
initial parton fractional momentum is quite large ($x_T\sim 0.6$ 
for $p_T=5$ GeV). The nucleon Fermi motion will cause the effective 
parton distribution to rise (EMC effect) and thus the modification
factor of hadron spectra also increase with $p_T$, even with jet
quenching.

For fixed $p_T$ and centrality in given $A+A$ collisions, one
can also use the energy dependence to study a non-Abelian feature 
of the parton energy loss, {\it i.e.}, the dependence on the 
color representation of the propagating parton. The energy loss 
for a gluon is 9/4 times larger than a quark. In this case,
one can exploit the well-known feature of the initial parton
distributions in nucleons (or nuclei) that quarks dominate at
large fractional momentum ($x$) while gluons dominate at small
$x$. Jet or large $p_T$ hadron production as a result of hard
scatterings of initial partons will be dominated by quarks at
large $x_T=2p_T/\sqrt{s}$ and by gluons at small $x_T$. Since
gluons lose 9/4 more energy than quarks, the energy dependence of
the large (but fixed) $p_T$ hadron spectra suppression due to
parton energy loss should reflect the transition from
quark-dominated jet production at low energy to gluon-dominated
jet production at high energy. Such sensitivity can be illustrated by
comparing hadron suppression factors with two different parton
energy losses: one for the QCD case where the energy loss for a
gluon is 9/4 times as large as that for a quark, {\it i.e.} $\Delta
E_g/\Delta E_q=9/4$; the other is for a non-QCD case where the
energy loss is chosen to be the same for both gluons and quarks.
Similarly, the average number of inelastic scatterings obeys
$\langle \frac{\Delta L}{\lambda}\rangle _g/ \langle \frac{\Delta
L}{\lambda}\rangle _q=9/4$ in the QCD case. For the non-QCD case
we are considering, the above ratio is set to one.
Shown in Fig.\ \ref{fig-edep} are the calculated $R_{AA}$ for 
neutral pions at fixed $p_T=6$ GeV in central $Au+Au$ collisions
as a function of $\sqrt{s}$ from 20 AGeV to 5500 AGeV
with both the QCD and non-QCD energy loss \cite{qwang}. 
In these calculations, the initial gluon number density and partons'
mean-free path is fixed to fit the overall hadron suppression in the most
central $Au+Au$ collisions at $\sqrt{s}=200$ GeV. For any other
energy, the initial gluon number is assumed 
to be proportional to the final measured total charged hadron
multiplicity per unit rapidity. One can see that due to the
dominant gluon bremsstrahlung or gluon energy loss at high energy
the $R_{AA}$ for the QCD case is more suppressed than the non-QCD
case where the gluon energy loss is assumed to be equal to
the quark. Such a unique energy dependence of
the high-$p_T$ hadron suppression can be tested by combining
$\sqrt{s}=200$ AGeV data with lower energy data or future data
from LHC experiments. As pointed out in Ref.~\cite{denterria}, the
existing data from SPS ($\sqrt{s}=17$ GeV) and RHIC ($\sqrt{s}=63$, 
130 and 200 GeV) already favor the case of non-Abelian energy loss,
which differs from the Abelian case by almost 50\% at $\sqrt{s}=200$ GeV.
The difference will grow to about a factor 2 at the LHC energy.

\section{Two-hadron correlations}

To reduce the sensitivity of jet quenching study to the underlying
jet spectra in the suppression of single hadron spectra, one can measure
two-hadron correlations at large $p_T$ \cite{star-jet}. It is accomplished by measuring
the spectra of hadrons associated with a triggered high $p_T$ hadron. Such
two-hadron correlation is effectively the ratio of dihadron and single hadron
spectra,
\begin{equation}
D_{AA}(z_T,\phi,p_T^{\rm trig})=p_T^{\rm trig}
\frac{d\sigma_{AA}/dp_T^{\rm asso}dp_T^{\rm trig}}
{d\sigma_{AA}/dp_T^{\rm trig}},
\label{corr}
\end{equation}
where $z_T=p_T^{\rm asso}/p_T^{\rm trig}$ and $\phi$ is the azimuthal angle
between the triggered and associated hadron. 

For $\phi<\pi//4$, the same-side two-hadron correlations are determined 
by the ratio of dihadron and single hadron
fragmentation functions. The dihadron fragmentation functions in terms of the
overlapping matrix between parton field operators and the final
hadron states have been defined and their DGLAP evolution equations 
have been derived recently \cite{majumder1}, which are similar to that of 
single hadron fragmentation functions. However, there are extra contributions
that are proportional to the convolution of two
single hadron fragmentation functions. These correspond to independent
fragmentation of both daughter partons after the parton split in the
radiative processes. Medium modification to the dihadron fragmentation functions
due to induced radiation was found to have the identical form as the DGLAP evolution 
equations \cite{majumder2}. These medium modifications depend on the 
same gluon correlation functions as in the modification to the 
single hadron fragmentation functions. Therefore, in the
numerical calculation of the medium modification of dihadron fragmentation
functions, there are no additional parameters involved. The predicted results
for jet quenching in DIS are in good agreement with 
HERMES data \cite{majumder2}. The nuclear modification is found to manifest 
mostly in the single hadron fragmentation functions. Since dihadron fragmentation 
functions already contain the information of single hadron fragmentation function, 
the modification to the remaining correlated distribution,
$D_q^{h_1h_2}(z_1,z_2)/D_q^{h_1}(z_1)$ 
is very small. This explains why the same-side two-hadron correlation
in central heavy-ion collisions remains approximately the same as 
in $p+p$ collisions \cite{star-jet}. However, trigger bias in heavy-ion collisions
could lead to some apparent change of dihadron correlations \cite{majumder2}.

For $\phi \sim \pi$, away-side two-hadron spectra are proportional to the 
product of two single fragmentation functions. Therefore, the away-side
two-hadron correlations in Eq.~(\ref{corr}) essentially reflect the medium
modification of the single fragmentation function in the opposite direction
of the triggered jet. For fixed $p_T^{\rm trig}$ and $p_T^{\rm asso}$, the
initial transverse momentum of the away-side jet can fluctuate. Such fluctuation
leads to a rather flat medium modification of the correlation function,
$I_AA(z_T)=D_{AA}(z_T)/D_{pp}(z_T)$ for large values of $z_T$ and it only 
starts to increase at small $z_T<0.2$ \cite{Wang:2003mm}. Recent measurements from
STAR have qualitatively confirmed this feature \cite{magestro}. For large
$p_T^{\rm trig}$, the correlation $I_{AA}(z_T)$ scales approximately with $z_T$.
It will be useful to verify this experimentally.

\section{Jet-induced collective excitation}

The suppression of high $p_T$ single inclusive hadron spectra and two-hadron
correlations can be attributed to modification of jet fragmentation in medium
via induced gluon bremsstrahlung. However, most of the studies have neglected the
interaction of the radiated gluons with the thermal medium. Such interaction
could potentially affect the spectra of soft hadrons associated with a jet.
Recent experimental studies of angular correlations of soft hadrons with respect
to a quenched jet indeed have revealed a peculiar pattern \cite{phenix-jet}. In
central $Au+Au$ collisions, soft hadrons associated with a quenched jet
(in the opposite direction of the triggered hadron) are peaked at a finite 
angle $\Delta\phi\sim 1$ away from the jet, whereas, they peak in the
direction of the jet in peripheral $Au+Au$ or $p+p$ collisions. 
This observation has lead to suggestions of different scenarios for the
interaction between soft partons and the thermal medium as the leading parton
propagates through the dense medium.

In the most simplified scenario, one can assume that the soft partons radiated
from the leading jet (or equivalently the recoil from elastic scattering) strongly
interact with the medium and are immediately thermalized. The energy deposited
by the jet through such strong interaction will then propagate through the
medium as a sound wave. Because the sound velocity $c_s$ is smaller than the
(light) velocity of the leading massless parton, a shock wave would eventually
develop \cite{stoecker}. The wave front of the shock will have a Mach cone 
angle $\cos \theta_M=c_s$. Such formation of shock wave or Mach cone has 
been demonstrated in hydrodynamical simulations \cite{shuryak}.

In the limit of slow thermalization of the radiated partons, a ring structure in
the final particle distribution can also be formed via Cerenkov gluon radiation
if the soft gluon can develop a space-like dispersion relation through interaction
with the thermal medium \cite{Ruppert}. Such space-like dispersion is possible
if the strongly interactive quark-gluon plasma has many colored resonant 
structures or partonic bound states. The transitional scattering between a soft
gluon and two bound states with different masses has been shown to lead to
a space-like dispersion for the soft gluon \cite{cerenkov1}. 
However, the total energy loss caused by Cerenkov gluon radiation is very small as 
compared to radiative energy loss induced by multiple parton scattering.
If one employs a space-like dispersion relation for soft gluons
in the calculation of induced gluon bremsstrahlung, the LPM interference will
produce a final gluon spectra with a peak at a finite angle \cite{cerenkov2}
\begin{equation}
\cos^2\theta_c=z+\frac{1-z}{\epsilon(\ell)},
\end{equation}
which is determined by gluon's dielectric constant $\epsilon(\ell)$, where
$z$ is the gluon's fractional energy. In the soft radiation limit $z\sim 0$, this 
corresponds exactly to the angle of classical Cerenkov radiation.
Such Cerenkov-like bremsstrahlung can induce a large energy loss.
For a large dielectric constant $\epsilon \gg 1+2/z^2LE$, the
corresponding total radiative parton energy loss is about twice that
from normal gluon bremsstrahlung \cite{cerenkov2}. The unique feature
of Cerenkov-like bremsstrahlung is that the Cerenkov cone size decreases
with the momentum of the soft gluon. On the other hand, Mach cone of
sonic shock wave is independent of the particle momentum.

\section{Summary}

In summary, the discovery and detailed study of jet quenching in high-energy
heavy-ion collisions at RHIC have provided strong evidence that a strongly
interactive quark-gluon plasma has been formed in the central $Au+Au$ collisions.
Such sQGP is opaque to energetic parton jets.  The initial gluon density
at $\tau_0=0.2$ fm/$c$ is estimated to be about 30 times higher than in
a cold nuclear matter and the energy density about 100 times higher. The peculiar
angular distribution of soft hadrons relative to the quenched jet could be an
indication of sonic shock wave or Cerenkov-like gluon bremsstrahlung. If confirmed
by further experimental test, this could provide another powerful information about
properties of the sQGP. 

Combined with many other aspects of jet quenching,
jet tomography has become a useful and powerful tool to study the properties
of dense matter in heavy-ion collisions.
With accumulation of data and development of new experimental analysis techniques,
the study of jet quenching via direct $\gamma$ tagged jet will become available.
Such study would be ideal because there will not be complicated trigger bias
effect as compared to high-$p_T$ hadron triggering. The suppression is expected
to be similar to the single hadron spectra \cite{gamma-jet}, 
except that one knows better the 
initial jet energy and therefore can study the energy dependence of the jet 
quenching. Two-hadron correlation in $\gamma$-tagged jet is also
better due to absence of correlated background because there is no 
trigger-biased correlation with the reaction plane.

\section*{Acknowledgement}

I would like to thank Abhjit Majumder and Qun Wang
for their collaboration on some of the work I reported in this talk.
This work was supported the Director, Office of Energy
Research, Office of High Energy and Nuclear Physics, Division of
Nuclear Physics, and by the Office of Basic Energy Science,
Division of Nuclear Science, of  the U.S. Department of Energy
under Contract No. DE-AC02-05CH11231.

\end{document}